\documentclass[letterpaper]{article}

\usepackage[T1]{fontenc}

\usepackage{geometry}
\usepackage{setspace}

\usepackage[super,sort&compress]{natbib}
\usepackage{graphicx}
\usepackage{subcaption}
\usepackage{multirow}
\usepackage{amsmath,amssymb,amsfonts}
\usepackage{amsthm}
\usepackage{mathrsfs}
\usepackage{xcolor}
\usepackage{textcomp}
\usepackage{booktabs}
\usepackage{comment}
\usepackage{tabularx}
\usepackage{array}
\usepackage{float}
\usepackage{placeins}
\usepackage{enumitem}
\usepackage{rotating} 

\usepackage{xurl}

\newcolumntype{L}[1]{>{\raggedright\arraybackslash}p{#1}}
\newcolumntype{Y}{>{\raggedright\arraybackslash}X}

\newcommand{\botrule}{\bottomrule}

\newcommand{\tablenote}[1]{\par\smallskip\footnotesize\noindent #1\par}

\usepackage{authblk}

\author[1,2,*]{Edoardo Rosci}
\author[1]{Giovanna Jona Lasinio}
\author[2]{Paola Michelozzi}
\author[2]{Massimo Stafoggia}

\affil[1]{Department of Statistical Sciences, Sapienza University of Rome,
Piazzale Aldo Moro 5, Rome, 00185, Italy}

\affil[2]{Department of Epidemiology, Lazio Region Health Service/ASL Roma 1,
Via Cristoforo Colombo 112, Rome, 00147, Italy}

\date{*Corresponding author: Edoardo Rosci ---
edoardo.rosci@uniroma1.it, e.rosci@deplazio.it}

\title{Bayesian spatio-temporal exposure modelling of air pollution in Rome, Italy, and short-term effects on cause-specific mortality}

\begin{document}

\maketitle

\begin{abstract}
Assessing fine-scale spatio-temporal air pollution contrasts in urban contexts is a major challenge for environmental epidemiology. We propose a Bayesian spatio-temporal model to predict daily concentrations of NO\textsubscript{2}, PM\textsubscript{10}, and PM\textsubscript{2.5} in Rome (Italy), over 2011--2022, on a fine grid scale (1~km), using data from 8 to 13 monitoring stations, depending on pollutant. The model includes meteorological and temporal (working day/weekend) fixed effects together with a lag-1 autoregressive spatio-temporal random effect aimed at capturing spatial and daily dependence. Predictive performance was assessed by leave-one-site-out cross-validation. Estimated exposures were then linked to geolocated cause-specific mortality data for Rome (2012--2019), and a case-crossover time-stratified approach was adopted to investigate acute effects. Cross-validation showed good overall predictive performance, with larger errors at high-traffic sites. Estimated exposure (mean lag~0--5) was positively associated with natural-cause mortality, with percent increase in risk of 1.3 (95\% CI: 0.6--2.0) for PM\textsubscript{10}, 2.1 for NO\textsubscript{2} (1.4--2.9), and 2.4 for PM\textsubscript{2.5} (1.4--3.3) per 10~$\mu$g/m$^3$ increase. Corresponding estimates when using the city-specific daily average exposure, instead of our 1~km resolution model, were of comparable magnitude. The proposed Bayesian spatio-temporal framework provides reliable fine-scale exposure estimates for epidemiological use, with results consistent across independent exposure estimates, supporting its application in urban air-pollution health studies.
\end{abstract}

\section*{Keywords}
Case-crossover, Environmental epidemiology, Markov chain Monte Carlo, Nitrogen dioxide, Particulate matter, Urban health

\section{Introduction}\label{EH::intro}
Exposure to air pollution in urban contexts represents a huge epidemiological challenge, because of the complexity of capturing small-scale contrasts over space and time. Among the many air pollutants, those of greatest epidemiological concern are nitrogen dioxide (NO\textsubscript{2}) and particulate matter with aerodynamic diameter $\leq$10~$\mu$m (PM\textsubscript{10}) and $\leq$2.5~
$\mu$m (PM\textsubscript{2.5}). As reviewed, for instance, by \citet{orellano}, the link between short-term exposure to these pollutants and all-cause and cause-specific mortality is now well established in the literature.
Concerning exposure assessment, several statistical models have been proposed for air pollution exposure. To give an example, Land-Use Regression models have been extensively applied to capture spatial contrasts for long-term exposure assessment, benefiting from the increasing availability of high-resolution spatial and spatio-temporal predictors \citep{xu_lur}. More recently, machine-learning approaches, such as the three-stage Random Forest (RF) model developed by \citet{stafoggia_2019} for PM\textsubscript{10} and PM\textsubscript{2.5} in Italy (2013--2015), have been used to obtain fine-scale (1~km) predictions. Hierarchical and state-space specifications, in which random effects capture spatio-temporal dependence, represent a possible alternative approach: \citet{dabass} estimated daily PM\textsubscript{2.5} exposure with a kriged spatio-temporal model and linked it to cardiovascular mortality in a case-crossover analysis, while \citet{pirani_blangiardo} compared several specifications, including an autoregressive one, for short-term PM\textsubscript{10} exposure assessment in London (UK).
Moving from the exposure assessment to the epidemiological side, model predictions can be linked to cohort data to investigate chronic effects, or to geolocated mortality or morbidity outcomes to investigate acute effects in a case-crossover design \citep{MaclureCaseCrossover}. These models have been widely applied in air-pollution and urban-health research, as described, for instance, in \citet{Tobias2024}.

In this framework, Rome (Italy) represents a relevant setting for this application: as one of the biggest cities in Europe by population and area, road traffic remains a major contributor to urban air pollution. Both PM\textsubscript{2.5} and NO\textsubscript{2} concentrations have shown an association with mortality, especially under high-traffic conditions \citep{michelozzi1998, cesaroni2013, Renzi2017}.

The aims of the present work are two. The first is to build and validate, through Bayesian inference, a spatio-temporal autoregressive model for daily air pollution exposure in Rome municipality between 2011 and 2022, and to construct daily fine-scale exposure surfaces. The second is to apply the estimated exposures in a case-crossover analysis of short-term air pollution-mortality associations in Rome. The proposed exposure model is hence evaluated in two ways: a purely statistical one, via cross-validation against held-out monitoring stations, and an epidemiologically comparative one, comparing case-crossover mortality associations obtained with the model-estimated exposure against those obtained with independent sources (station averages and the RF estimates of \citet{stafoggia_2019}).

\section{Methods}\label{EH::methods}
\subsection{Study area and data}
This study focuses on the Rome municipality, the capital of the Lazio Region and the most populous municipality in Italy, with 2,747,290 registered residents as of 31 December 2024 \citep{istat2024}. The municipal area covers approximately 1,285~km$^2$, making Rome one of the largest cities in Europe by extent. As shown on the left side of Figure~\ref{plot::mappa_staz}, Rome is located in central Italy, approximately 25~km from the Tyrrhenian coast, at coordinates $41.9^{\circ}$ N, $12.5^{\circ}$ E.

\subsubsection{Air pollution data}
\begin{figure}[H]
  \centering
  \includegraphics[width=0.85\linewidth,trim=10 40 10 40,clip]{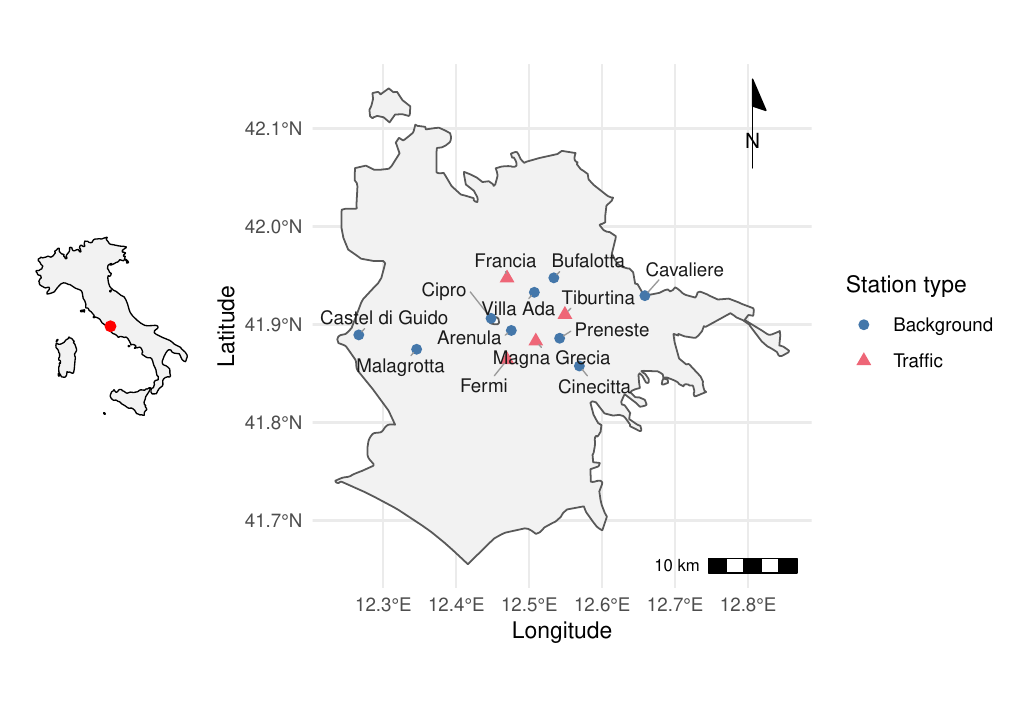}
 \caption{Rome municipality location within the Italian peninsula (left). Total sample of considered monitoring stations locations within the Rome municipality, coloured by aggregated station type (Background/Traffic) (right).}
  \label{plot::mappa_staz}
\end{figure}

Air pollution data for NO\textsubscript{2}, PM\textsubscript{10} and PM\textsubscript{2.5} were derived from the Regional Agency for Environmental Protection of Lazio (ARPA Lazio) monitoring network. Original hourly values (in $\mu$g/m$^3$) between 1 January 2011 and 31 December 2022 were summarised as daily averages. Days with more than six missing hourly detections were considered missing: therefore, restricting the analysis to stations with less than 5\% overall missingness resulted in 13 NO\textsubscript{2} stations, 12 PM\textsubscript{10} stations, and 8 PM\textsubscript{2.5} stations. A more detailed description of the stations is provided in Table~\ref{tab::staz}. Moreover, as represented in the right side of Figure~\ref{plot::mappa_staz}, monitoring stations were concentrated mainly in the centre of the municipality. Additional descriptive plots of observed pollutant concentrations are provided in the Supporting Information.

\begin{sidewaystable}[!htbp]
\centering
\caption{Summary of Rome daily concentrations by monitoring station and pollutant.}
\label{tab::staz}
\footnotesize
\setlength{\tabcolsep}{3pt}
\renewcommand{\arraystretch}{1.00}
\begin{tabular}{L{1.8cm} L{2.2cm} L{1.4cm} c c c *{3}{>{\centering\arraybackslash}p{1.6cm} >{\centering\arraybackslash}p{0.9cm}}}
\toprule
\multirow{2}{*}{\textbf{Station}} & \multirow{2}{*}{\textbf{Station Type}} & \multirow{2}{*}{\textbf{Area Type}} & \multirow{2}{*}{\textbf{Lat}} & \multirow{2}{*}{\textbf{Lon}} & \multirow{2}{*}{\textbf{Period}} &
\multicolumn{2}{c}{\textbf{NO$_{2}$}} & \multicolumn{2}{c}{\textbf{PM$_{10}$}} & \multicolumn{2}{c}{\textbf{PM$_{2.5}$}} \\
\cmidrule(lr){7-8} \cmidrule(lr){9-10} \cmidrule(lr){11-12}
 & & & & & & \textbf{Median (IQR)} & \textbf{NA\%} & \textbf{Median (IQR)} & \textbf{NA\%} & \textbf{Median (IQR)} & \textbf{NA\%} \\
\midrule
CDG & Background Rural    & Rural    & 41.8895 & 12.2663 & 2011--2022 & 10.6 (6.7--17.9)  & 4.4 & 19.0 (15.0--25.0) & 2.3 & 11.0 (8.0--15.0)  & 2.3 \\
CAV & Background Suburban & Suburban & 41.9295 & 12.6585 & 2011--2022 & 25.8 (17.9--36.9) & 3.0 & 22.0 (17.0--30.0) & 3.1 & 13.0 (9.0--19.0)  & 3.4 \\
MAL & Background Suburban & Suburban & 41.8748 & 12.3456 & 2011--2022 & 18.2 (11.5--28.3) & 3.5 & 21.0 (16.0--28.0) & 2.9 & 12.0 (9.0--18.0)  & 3.2 \\
ARE & Background Urban    & Urban    & 41.8940 & 12.4754 & 2011--2022 & 42.0 (30.9--55.2) & 1.9 & 24.0 (18.0--32.0) & 3.7 & 13.0 (9.0--19.0)  & 3.0 \\
BUF & Background Urban    & Urban    & 41.9477 & 12.5337 & 2011--2022 & 34.1 (24.7--45.9) & 3.4 & 25.0 (18.0--33.0) & 2.5 & --                  & --  \\
CIN & Background Urban    & Urban    & 41.8577 & 12.5687 & 2011--2022 & 35.3 (24.8--48.7) & 2.2 & 26.0 (20.0--35.0) & 4.5 & 15.0 (10.0--21.0) & 4.9 \\
CIP & Background Urban    & Urban    & 41.9064 & 12.4476 & 2011--2022 & 41.9 (30.3--55.2) & 1.6 & 23.0 (17.0--31.0) & 1.6 & 13.0 (9.0--18.0)  & 3.7 \\
PRE & Background Urban    & Urban    & 41.8860 & 12.5416 & 2011--2022 & 37.1 (25.6--50.2) & 1.5 & --                  & --  & --                  & --  \\
VAD & Background Urban    & Urban    & 41.9329 & 12.5069 & 2011--2022 & 28.8 (21.3--40.3) & 3.4 & 22.0 (17.0--28.0) & 2.8 & 12.0 (9.0--18.0)  & 4.6 \\
FER & Traffic Urban       & Urban    & 41.8640 & 12.4696 & 2011--2022 & 60.0 (47.6--72.9) & 2.6 & 28.0 (22.0--36.0) & 3.0 & --                  & --  \\
FRA & Traffic Urban       & Urban    & 41.9474 & 12.4696 & 2011--2022 & 56.7 (43.8--68.9) & 2.1 & 27.0 (21.0--35.0) & 1.5 & 15.0 (11.0--21.0) & 1.7 \\
MGR & Traffic Urban       & Urban    & 41.8831 & 12.5090 & 2011--2022 & 55.4 (41.1--69.5) & 3.9 & 25.0 (19.0--33.0) & 2.3 & --                  & --  \\
TIB & Traffic Urban       & Urban    & 41.9103 & 12.5489 & 2011--2022 & 49.2 (37.3--63.0) & 1.5 & 29.0 (22.0--38.0) & 1.3 & --                  & --  \\
\botrule
\end{tabular}
\tablenote{Values expressed in $\mu$g/m$^3$. Median (IQR) computed over daily mean values, 2011--2022. NA\% denotes the percentage of missing daily observations over the same period relative to the expected number of daily records. ``--'' indicates the pollutant is not monitored at that station. Station codes: ARE = Arenula; BUF = Bufalotta; CAV = Cavaliere; CDG = Castel di Guido; CIN = Cinecitt\`a; CIP = Cipro; FER = Fermi; FRA = Francia; MAL = Malagrotta; MGR = Magna Grecia; PRE = Preneste; TIB = Tiburtina; VAD = Villa Ada.}
\end{sidewaystable}

\subsubsection{Covariates}
Meteorological and climatic variables included in the analysis as external covariates were derived from Copernicus Climate Data Store using the ERA5 and ERA5-Land reanalyses \citep{era5_cds, era5land_cds}. Hourly values were aggregated to daily summaries and assigned to each monitor by nearest-neighbour matching by space and date. Wind speed was derived from the respective zonal and meridional 10m wind components. A temporal dummy indicator (\textit{day\_we}) was also considered to account for the effect of weekends and holidays. A full list of the included covariates, with their respective definitions, is reported in Table~\ref{tab::cov}.

\begin{sidewaystable}[!htbp]
\centering
\caption{Summary of temporal and spatio-temporal covariates included in model selection.}
\label{tab::cov}
\footnotesize
\setlength{\tabcolsep}{3pt}
\renewcommand{\arraystretch}{1.00}
\begin{tabular}{L{3.1cm} L{1.2cm} L{1.6cm} L{1.5cm} L{1.5cm} L{1.9cm} L{1.8cm} L{3.2cm}}
\toprule
\textbf{Full name} &
\textbf{Short name} &
\textbf{Type} &
\textbf{Source} &
\textbf{Unit} &
\textbf{Spatial resolution} &
\textbf{Temporal coverage} &
\textbf{Daily aggregation} \\
\midrule
2m Temperature & \texttt{t2m} & Spatio-temporal & ERA5-Land & $^\circ$C & 0.1$^\circ$ & 2011--2022 & Daily mean \\
2m Dew-point temperature & \texttt{d2m} & Spatio-temporal & ERA5-Land & $^\circ$C & 0.1$^\circ$ & 2011--2022 & Daily mean \\
Total precipitation & \texttt{tp} & Spatio-temporal & ERA5-Land & mm & 0.1$^\circ$ & 2011--2022 & Daily sum \\
Surface pressure & \texttt{sp} & Spatio-temporal & ERA5-Land & hPa & 0.1$^\circ$ & 2011--2022 & Daily mean \\
10m wind speed & \texttt{ws} & Spatio-temporal & ERA5-Land & $\mathrm{m\,s^{-1}}$ & 0.1$^\circ$ & 2011--2022 & Daily mean \\
Boundary layer height & \texttt{blh} & Spatio-temporal & ERA5 & m & 0.25$^\circ$ & 2011--2022 & Value at 00:00; \newline Value at 12:00 \\
\addlinespace
Weekend/holidays effect & \texttt{day\_we} & Temporal & -- & 0/1 & -- & 2011--2022 & Dummy indicator \newline (=1 in WE/holidays) \\
\botrule
\end{tabular}
\tablenote{\textit{2m Temperature} is the temperature of the air at 2m above the surface of land, sea or inland waters. \textit{2m Dew-point} represents the temperature to which the air, at 2 metres above the surface of the Earth, would have to be cooled for saturation to occur. \textit{Total precipitation} is the accumulated liquid and frozen water, including rain and snow, that falls to the Earth's surface. \textit{Surface pressure} represents the pressure of the atmosphere on the surface of land, sea and inland water. \textit{10m wind speed} was derived from the respective zonal (eastward, horizontal speed of air moving towards the east at a height of ten metres above the surface of the Earth) and meridional (northward, horizontal speed of air moving towards the north at a height of ten metres above the surface of the Earth) 10m wind components. \textit{Boundary layer height} is the depth of air next to the Earth's surface which is most affected by the resistance to the transfer of momentum, heat or moisture across the surface. \citep{era5_cds, era5land_cds}}
\end{sidewaystable}

\subsubsection{Mortality data}
Cause-specific geolocated mortality data were available for natural (by International Classification of Diseases, ICD: ICD-9 1--799, ICD-10 A00--R99), cardiovascular (ICD-9 390--459, ICD-10 I00--I99), and respiratory (ICD-9 460--519, ICD-10 J00--J99) causes between 2012 and 2019. Individual records were retrieved from the Rome Longitudinal Study, an administrative cohort encompassing the entire population of Rome aged 30 years or older at baseline. The cohort was enrolled at the Census date of Oct. 9, 2011, and followed up until 2019. For all subjects, geocodes of the residential addresses at baseline were available, plus potential changes of address during follow-up. For the purposes of this study, we considered only deceased individuals, and their geocoded residential addresses at the time of death were linked to the nearest predicted grid cell and matched by date. Each record also included information on age and sex. Table~\ref{tab:mortality_descriptive} summarises the distribution of deaths by cause, sex, age group, and season. Additional descriptive material on the spatial and temporal distribution of death events is provided in the Supporting Information.

\begin{table}[!htbp]
\caption{Descriptive characteristics of deaths included in the case-crossover analysis, Rome, 2012--2019.}
\label{tab:mortality_descriptive}
\centering
\begin{tabular}{lccc}
\toprule
 & \textbf{Natural} & \textbf{Cardiovascular} & \textbf{Respiratory} \\
\midrule
N deaths, total & 188,428 & 65,177 & 14,233 \\
\addlinespace
\textit{Sex} & & & \\
\quad Male   & 87,830 (46.6\%)  & 27,868 (42.8\%) & 6,689 (47.0\%) \\
\quad Female & 100,598 (53.4\%) & 37,309 (57.2\%) & 7,544 (53.0\%) \\
\addlinespace
\textit{Age group} & & & \\
\quad 30--64 & 16,860 (8.9\%)  & 3,218 (4.9\%)   & 449 (3.2\%)    \\
\quad 65--74 & 24,757 (13.1\%) & 5,294 (8.1\%)   & 1,191 (8.4\%)  \\
\quad 75--84 & 56,421 (29.9\%) & 16,485 (25.3\%) & 4,054 (28.5\%) \\
\quad 85+    & 90,390 (48.0\%) & 40,180 (61.6\%) & 8,539 (60.0\%) \\
\addlinespace
\textit{Season} & & & \\
\quad Winter & 52,915 (28.1\%) & 19,239 (29.5\%) & 4,647 (32.6\%) \\
\quad Spring & 46,264 (24.6\%) & 16,042 (24.6\%) & 3,655 (25.7\%) \\
\quad Summer & 45,128 (23.9\%) & 15,094 (23.2\%) & 3,051 (21.4\%) \\
\quad Autumn & 44,121 (23.4\%) & 14,802 (22.7\%) & 2,880 (20.2\%) \\
\botrule
\end{tabular}
\tablenote{Cardiovascular and respiratory deaths are subsets of natural-cause deaths, not mutually exclusive categories.}
\end{table}

\clearpage

\subsection{Exposure model}
In the spirit of \citet{jorge_2022}, the Bayesian spatio-temporal autoregressive model proposed in this work utilises a continuous space specification, keeping the temporal one in two separated discrete scales (specifically, years and days within years).

\subsubsection{Model specification}
Let $\mathbf{s} \in D$ denote a generic spatial location, with $D$ representing the whole Rome municipality. Even though $D$ denotes the entire spatial domain, pollution concentrations are actually observed at the $n$ monitoring sites located at points $\mathbf{s}_i \in \{\mathbf{s}_1, \dots, \mathbf{s}_n\} \subset D$. For each day $\ell = 1,\dots,365$ (29 February was excluded from leap years) of each year $t = 1,\dots,12$ (representing 2011--2022), let
\begin{equation}
    \mathbf{Y}_{t\ell}^{(j)} = \left(Y_{t\ell}^{(j)}(\mathbf{s}_1), \dots, Y_{t\ell}^{(j)}(\mathbf{s}_n)\right)^{\top}
\end{equation}
denote the vector of transformed concentrations of pollutant $(j)$, obtained by stacking observations across the $n$ monitoring sites. Original pollution concentrations were transformed to reduce skewness and heavy tails, as in similar air pollution modelling applications \citep{Fasso2007}: a square-root transformation was selected for NO\textsubscript{2} and a $\log(1+x)$ transformation for PMs (more details in the Supporting Information). Moreover, models were fitted for each pollutant separately, hence in the following we will omit the superscript $(j)$. Let $\mathbf{X}_{t\ell}$ denote the $n \times k$ matrix of temporal and spatio-temporal covariates for day $\ell$ of year $t$, whose $i$-th row contains the covariates observed at location $\mathbf{s}_i$, and where $k$ is the number of predictors selected in the best model specification. \\
The model can then be formalised as:
\begin{align}
    \label{mod_eq}
    \mathbf{Y}_{t\ell} &= \mathbf{X}_{t\ell}\boldsymbol{\beta} + \mathbf{w}_{t\ell}+\boldsymbol{\varepsilon}_{t\ell}, \\
    \mathbf{w}_{t\ell} &= \rho_w \mathbf{w}_{t,\ell-1} + \boldsymbol{\delta_{t\ell}}, ~ \boldsymbol{\delta}_{t\ell} \sim N_n\left(\boldsymbol{0}_n, \sigma^2_w\mathbf{R}_{\phi_w}\right), \\
    \mathbf{w}_{t1} &\sim N_n\left(\boldsymbol{0}_n, \frac{\sigma^2_w}{1-\rho^2_w} \mathbf{R}_{\phi_w}\right), \\
    \boldsymbol{\varepsilon}_{t\ell} &\sim N_n\left(\boldsymbol{0}_n, \sigma^2_{\varepsilon} \mathbf{I}_n\right),
\end{align}
with $\boldsymbol{\delta}_{t\ell}$, $\mathbf{w}_{t1}$, and $\boldsymbol{\varepsilon}_{t\ell}$ assumed mutually independent across time indices and independent of each other.
The term $\mathbf{X}_{t\ell}\boldsymbol{\beta}$ represents the effect of temporal and spatio-temporal fixed effects. The random effect $\mathbf{w}_{t\ell}$ captures spatio-temporal dependence, thanks to an autoregressive structure of order 1: hence, in this case $\rho_w$ governs the strength of the temporal autocorrelation. The spatial dependence of the random effect is modelled with scale parameter $\sigma^2_w$ and decay $\phi_w$, with $\mathbf{R}_{\phi_w} = \exp(-\phi_w \ d(\mathbf{s}_i, \mathbf{s}_j)), ~ \mathbf{s}_i, \mathbf{s}_j \in \left\{\mathbf{s}_1, \dots, \mathbf{s}_n\right\}$ representing an exponential correlation function. Lastly, $\boldsymbol{\varepsilon}_{t\ell}$ represents the global error term, with variance parameter $\sigma_\varepsilon^2$.

\subsubsection{Model inference and covariate selection}
Model inference was conducted within a Bayesian Markov chain Monte Carlo \cite{Gelfand1990} (MCMC) framework, with the \texttt{spTReg} R package (version 0.0.0.17, \cite{sptreg}) performing a Metropolis-within-Gibbs algorithm to obtain samples from the joint posterior distribution. Additional information regarding MCMC priors, posterior convergence, computational times, and Bayesian kriging is included in the Supporting Information.
Before fitting final MCMC specifications, covariate selection, restricted to the temporal and spatio-temporal fixed effects, was conducted in a frequentist framework. Predictors were first screened considering generalised Variance Inflation Factor  \cite{vif} (GVIF) to reduce multicollinearity. Final selected covariates by Bayesian Information Criterion \cite{SchwarzBIC} (BIC) are summarised in Table~\ref{tab:best_mod}.

\begin{table}[!htbp]
\centering
\caption{Covariates included in the best model specification for each pollutant according to the BIC.}
\label{tab:best_mod}
\begin{tabular}{lcc}
\toprule
\textbf{Pollutant} & \textbf{N. covariates} & \textbf{Selected covariates} \\
\midrule
NO$_2$    & 6 & day\_we, blh\_12, d2m, t2m, tp, sp \\
PM$_{10}$ & 7 & day\_we, blh\_12, d2m, t2m, tp, sp, ws \\
PM$_{2.5}$ & 6 & day\_we, blh\_12, d2m, t2m, sp, ws \\
\botrule
\end{tabular}
\tablenote{Abbreviations: day\_we = weekend/holidays dummy indicator; blh\_12 = boundary layer height at 12:00; d2m = 2m dew-point temperature; t2m = 2m temperature; tp = total precipitation; sp = surface pressure; ws = 10m wind speed. \\Full covariate definitions are reported in Table~\ref{tab::cov}.}
\end{table}

In the Supporting Information we assessed the robustness of the frequentist screening by repeating the process under a Bayesian paradigm, using comparable specifications. 

\subsubsection{Cross-validation}
Models were evaluated in terms of predictive performance by a leave-one-site-out (LOSO) cross-validation (CV): for each pollutant and iteration, the model was trained on $n-1$ stations, using the $n$-th as test set. As metrics for evaluation, we report Root Mean Square Error (RMSE) and Pearson correlation (r) between observed and predicted values.
Additional metrics are provided in the Supporting Information, together with in-sample predictive performance as a general goodness-of-fit check.

\subsection{Case-crossover}
The association between short-term air pollution exposure and mortality was assessed using a case-crossover design \citep{MaclureCaseCrossover}. Among the referent selection strategies available for case-crossover studies, we adopted the time-stratified approach \citep{JanesCaseCrossover}, in which calendar time is partitioned into fixed strata defined by year, month, and day of the week; for each case, all other days falling within the same stratum as the event day serve as control (referent) days. The posterior median predicted concentration for each grid cell and date was used as the exposure value in the primary analysis.

\subsubsection{Statistical model}
Case-crossover data were analysed using conditional logistic regression, conditioning on the matched stratum defined previously (implemented via the \textit{clogit} function in the \texttt{survival} R package \cite{survival}, version 3.7.0). Let $m = 1, \dots, M$ index case-specific strata, one per case: control days entering stratum $m$ are selected by the calendar rule described above. Let $y_{im}$ denote the case indicator for day $i$ within stratum $m$ ($y_{im}=1$ for the case day, $0$ for each matched control day). The model can be written as
\begin{align}
    \text{logit}\left[\Pr(y_{im}=1)\right] &= \alpha_m + \eta_{im}, \\
    \eta_{im} &= f(z_{im}) + \text{cb}(T_{im}, u; \boldsymbol{\theta}), \qquad z_{im} = \frac{1}{6}\sum_{k=0}^{5} x_{i-k,m},
\end{align}
where $\alpha_m$ is a stratum-specific intercept absorbing all time-invariant characteristics of the stratum (and, by design, of the matched individual); $x_{i-k,m}$ is the estimated pollutant concentration $k$ days before day $i$, so that $z_{im}$ is the mean lag~0--5 exposure kept as reference specification; $f(\cdot)$ is linear in the primary analysis and modelled via a natural cubic spline (2 degrees of freedom) in the exposure-response analysis. We use $\text{cb}(T_{im}, u; \boldsymbol{\theta})$ to denote the distributed lag nonlinear (DLNM) cross-basis \citep{GasparriniDLNM} for daily mean temperature (2m Temperature, derived from ERA5-Land, as previously described in Table~\ref{tab::cov}) over lags $u = 0, \dots, 14$ days, with $\boldsymbol{\theta}$ the vector of coefficients associated with the cross-basis functions. The cross-basis was implemented using the \texttt{dlnm} R package (version 2.4.10, \citep{GasparriniDLNMpackage}): the exposure-response dimension is modelled with a natural cubic spline (3 degrees of freedom), and the lag-response dimension is modelled with a natural cubic spline with internal knots at lag~1, 3, and 7 days.
The choice of the mean lag~0--5 window was informed by diagnostic analyses on natural-cause mortality (reported and discussed in the Supporting Information), and extended to cardiovascular and respiratory causes for comparability. For the linear specification, association results are estimated as rate ratios ($\text{RR} = \exp(10\beta)$) and then converted into percent increase in risk ($\text{IR\%} = 100(\text{RR}-1)$) per 10 $\mu$g/m$^3$ increase in exposure, with 95\% confidence intervals (CI). Additional tools for describing case-crossover datasets are reported in the Supporting Information.

\subsubsection{Sensitivity comparison with other exposure sources}
To assess robustness, case-crossover results based on the model-estimated exposure were compared against two benchmark sources: the daily average of observed concentrations across all municipal monitoring stations, for each pollutant (station mean), and the RF estimates of \citet{stafoggia_2019} (PM\textsubscript{10} and PM\textsubscript{2.5} only, spatially resolved at grid-cell level). Both sources were linked to the case-crossover records using the same date (and, for the RF estimates, grid-cell) as the model-estimated exposure, and analysed with the same case-crossover structure described above. Additional descriptive summaries are provided in the Supporting Information.

\clearpage

\section{Results}\label{EH::results}

\subsection{Exposure model}
\label{subs::exp_res}
\subsubsection{Parameter estimates}
Table~\ref{tab:params} reports the posterior median and 95\% credible interval (CrI) of the intercept and other hyperparameters of the spatio-temporal model for each pollutant. The temporal parameter $\rho_w$ was high across pollutants (0.981--0.984), indicating strong daily auto-correlation of the
spatio-temporal random effect. The spatial decay $\phi_w$ was about one order of magnitude larger for NO$_2$ than for PM$_{10}$ and PM$_{2.5}$.

\begin{table}[!htbp]
\centering
\caption{Posterior estimates of the intercept and models' hyperparameters.}
\label{tab:params}
\footnotesize
\setlength{\tabcolsep}{6pt}
\renewcommand{\arraystretch}{1.15}
\begin{tabular}{l ccc}
\toprule
\textbf{Parameter} & \textbf{NO$_2$} & \textbf{PM$_{10}$} & \textbf{PM$_{2.5}$} \\
\midrule
Intercept            & 5.79 [5.17, 6.18]          & 3.20 [2.87, 3.40]          & 2.53 [2.29, 2.90]          \\
$\rho_w$             & 0.981 [0.979, 0.983]       & 0.984 [0.982, 0.986]       & 0.982 [0.978, 0.985]       \\
$\sigma^2_w$           & 0.771 [0.758, 0.787]       & 0.300 [0.294, 0.306]       & 0.314 [0.307, 0.321]       \\
$\phi_w$               & 0.0150 [0.0142, 0.0159]    & 0.00161 [0.00143, 0.00180] & 0.00110 [0.00092, 0.00133] \\
$\sigma^2_\varepsilon$ & 0.287 [0.283, 0.291]       & 0.125 [0.124, 0.126]       & 0.158 [0.156, 0.159]       \\
\botrule
\end{tabular}
\tablenote{Values expressed as (median [95\% CrI]). Values must be referred to transformed response variables (square-root for NO\textsubscript{2}, $\log(1+x)$ for PM\textsubscript{10} and PM\textsubscript{2.5}).}
\end{table}

\begin{figure}[H]
  \centering
  \includegraphics[width=0.7\linewidth]{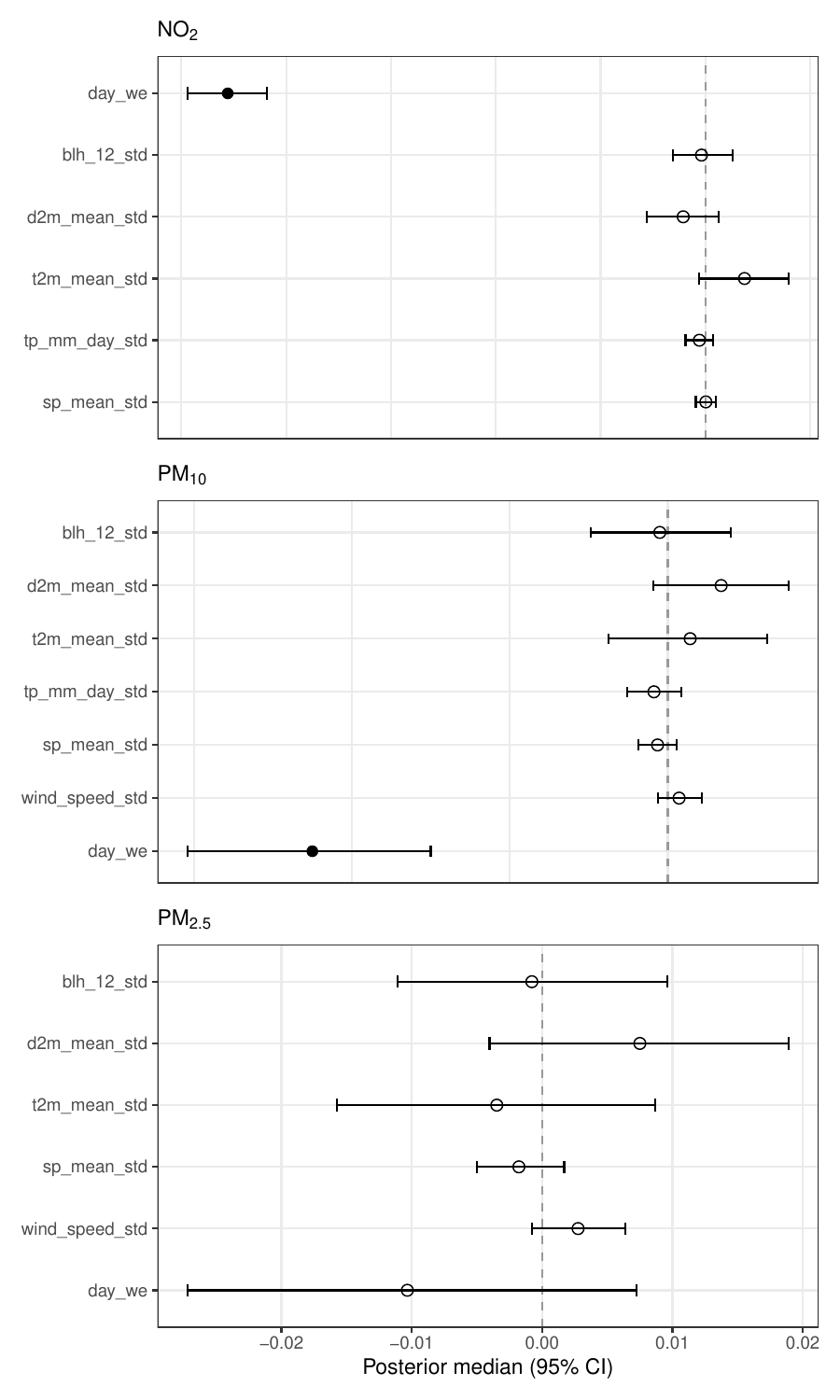}
  \caption{Posterior median and 95\% credible interval (CrI) of fixed effects (intercept excluded) for each pollutant: NO\textsubscript{2} (top), PM\textsubscript{10} (centre), and PM\textsubscript{2.5} (bottom). Filled circles indicate parameters whose 95\% CrI excludes zero; open circles indicate parameters whose interval includes zero. Values must be referred to transformed response variables (square-root for NO\textsubscript{2}, $\log(1+x)$ for PM\textsubscript{10} and PM\textsubscript{2.5}), and to standardised covariates using mean and standard deviation of the Rome prediction grid.}
  \label{fig:forest_beta}
\end{figure}

Fixed-effect estimates (excluding the intercept) are shown in
Figure~\ref{fig:forest_beta}. Weekend/holidays indicator was the only covariate whose 95\% CrI excluded zero for NO$_2$ and PM$_{10}$
(negative effect, i.e.\ lower concentrations on weekends), while for PM$_{2.5}$ its interval included zero. None of the meteorological covariates selected had a 95\% CrI excluding zero for any of the three pollutants. 

\subsubsection{Exposure surfaces}
\begin{figure}[H]
  \centering
  \includegraphics[width=\linewidth]{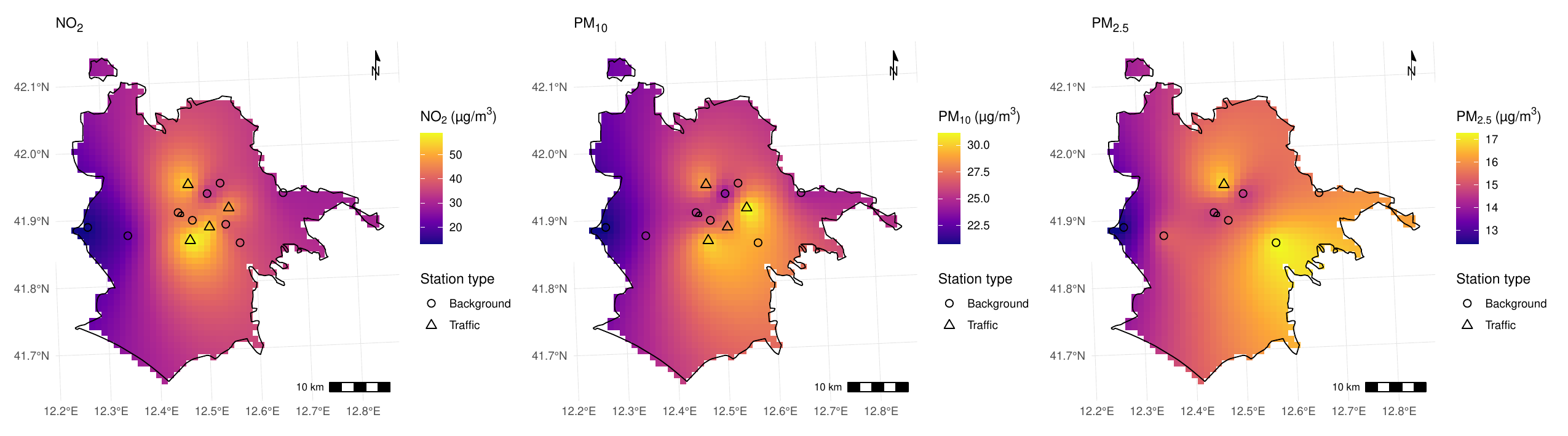}
  \caption{Estimated mean exposure surfaces for NO\textsubscript{2} (left), PM\textsubscript{10} (centre), and PM\textsubscript{2.5} (right). Shape by monitoring stations aggregated station type (Background/Traffic). Values expressed in $\mu$g/m$^3$.}
  \label{fig:exposure_surfaces}
\end{figure}

Figure~\ref{fig:exposure_surfaces} shows global averages of the estimated posterior daily medians over the whole 2011--2022 period, for each pollutant. NO\textsubscript{2} showed the greatest micro-scale variability, with local peaks around the traffic monitoring sites (especially Francia and Fermi). PM\textsubscript{10} showed its highest peak around the Tiburtina station. Lastly, PM\textsubscript{2.5} reflected the most uniform surface overall. More results, including posterior uncertainty quantification and a trend comparison of average surfaces (2012~v~2019), are included in the Supporting Information.

\subsubsection{Cross-validation results}
Table~\ref{tab::loso_main} reports an aggregated summary of the LOSO-CV results, showing both metrics (r and RMSE in the original $\mu$g/m$^3$ scale). Overall, predictive performance was good across pollutants and for both aggregated station types (Background/Traffic). NO\textsubscript{2} at traffic sites showed the lowest correlation and highest RMSE, whilst PM\textsubscript{10} and PM\textsubscript{2.5} had higher correlations and generally lower RMSEs. Full station-wise LOSO-CV results with more metrics are reported in the Supporting Information.

\begin{table}[!htbp]
\centering
\caption{Leave-one-site-out cross-validation metrics, aggregated by station type and pollutant.}
\label{tab::loso_main}
\footnotesize
\setlength{\tabcolsep}{6pt}
\renewcommand{\arraystretch}{1.15}

\begin{tabular}{llcc}
\toprule
\textbf{Pollutant} & \textbf{Station type} & \textbf{r} & \textbf{RMSE} \\
\midrule
\multirow{2}{*}{NO$_2$}
  & Background & 0.849 (0.720--0.930) & 10.9 (6.4--17.6) \\
  & Traffic    & 0.835 (0.706--0.898) & 20.2 (14.6--27.9) \\
\addlinespace

\multirow{2}{*}{PM$_{10}$}
  & Background & 0.913 (0.819--0.951) & 5.8 (3.8--8.2) \\
  & Traffic    & 0.921 (0.893--0.939) & 6.9 (4.6--8.5) \\
\addlinespace

\multirow{2}{*}{PM$_{2.5}$}
  & Background & 0.931 (0.895--0.957) & 4.2 (2.7--5.5) \\
  & Traffic    & 0.931\textsuperscript{a} & 4.8\textsuperscript{a} \\
\botrule
\end{tabular}

\tablenote{Values are expressed as mean (range) across stations of each type. RMSE is expressed in $\mu$g/m$^3$.
Number of stations per group: Background N = 9, 8, 7 for NO$_2$, PM$_{10}$ and PM$_{2.5}$, respectively;
Traffic N = 4, 4, 1.
\textsuperscript{a}Single traffic station monitoring PM$_{2.5}$ (Francia); range not applicable.}
\end{table}

\newpage

\subsection{Case-crossover results}
\label{subs:cc_res}

In Table~\ref{tab:cc_main} we report estimated IR\% per 10~$\mu$g/m$^3$ increase in exposure, for the lag~0--5 linear model for natural causes, using the model-estimated exposure. NO\textsubscript{2} and PM\textsubscript{2.5} showed the largest increases in risk using this scale (in both cases IR values higher than 2\%), whilst PM\textsubscript{10} remained below 1.3\%. In all cases the 95\% intervals excluded 0. To give a brief comparison with station mean independent source for natural cause mortality, resulting estimated IRs were 2.45 (1.56--3.35) for NO\textsubscript{2}, 1.33 (0.58--2.07) for PM\textsubscript{10}, and 2.37 (1.37--3.39) for PM\textsubscript{2.5}. Full results including all independent exposure sources considered are available in Supplementary Information.

\begin{table}[!htbp]
\caption{Association between short-term exposure to air pollutants (lag~0--5 mean, model-estimated exposure) and natural-cause mortality, Rome 2012--2019.}
\label{tab:cc_main}
\centering
\begin{tabular*}{0.5\textwidth}{@{\extracolsep{\fill}}lc@{}}
\toprule
\textbf{Pollutant} & \textbf{IR (95\% CI)} \\
\midrule
NO$_2$      & 2.14 (1.36--2.93) \\
PM$_{10}$   & 1.29 (0.59--2.00)\\
PM$_{2.5}$  & 2.35 (1.39--3.31) \\
\botrule
\end{tabular*}
\tablenote{IR~(\%) = percent increase in risk, $100\times(\mathrm{RR}-1)$ per 10~$\mu$g/m$^3$ increase in exposure (mean lag~0--5), assuming a linear exposure--response. Estimates based on the model-estimated exposure source.}
\end{table}

To assess whether a linear exposure-response relationship properly described these associations, we compared the linear specification with a natural-spline model (2 df) on the same lag~0--5 mean exposure using a likelihood ratio test (LRT), and the Akaike Information Criterion \cite{AIC} (AIC). For PM$_{10}$ and PM$_{2.5}$ the linear model was not significantly improved (LRT p = 0.21 and p = 0.76, respectively). For NO$_2$ instead, the flexible model fit significantly better (LRT p = 0.02) than the linear specification. The full LRT and $\Delta$AIC results for all pollutants and data sources are reported in the Supporting Information.

\begin{figure}[H]
  \centering
  \includegraphics[width=\linewidth]{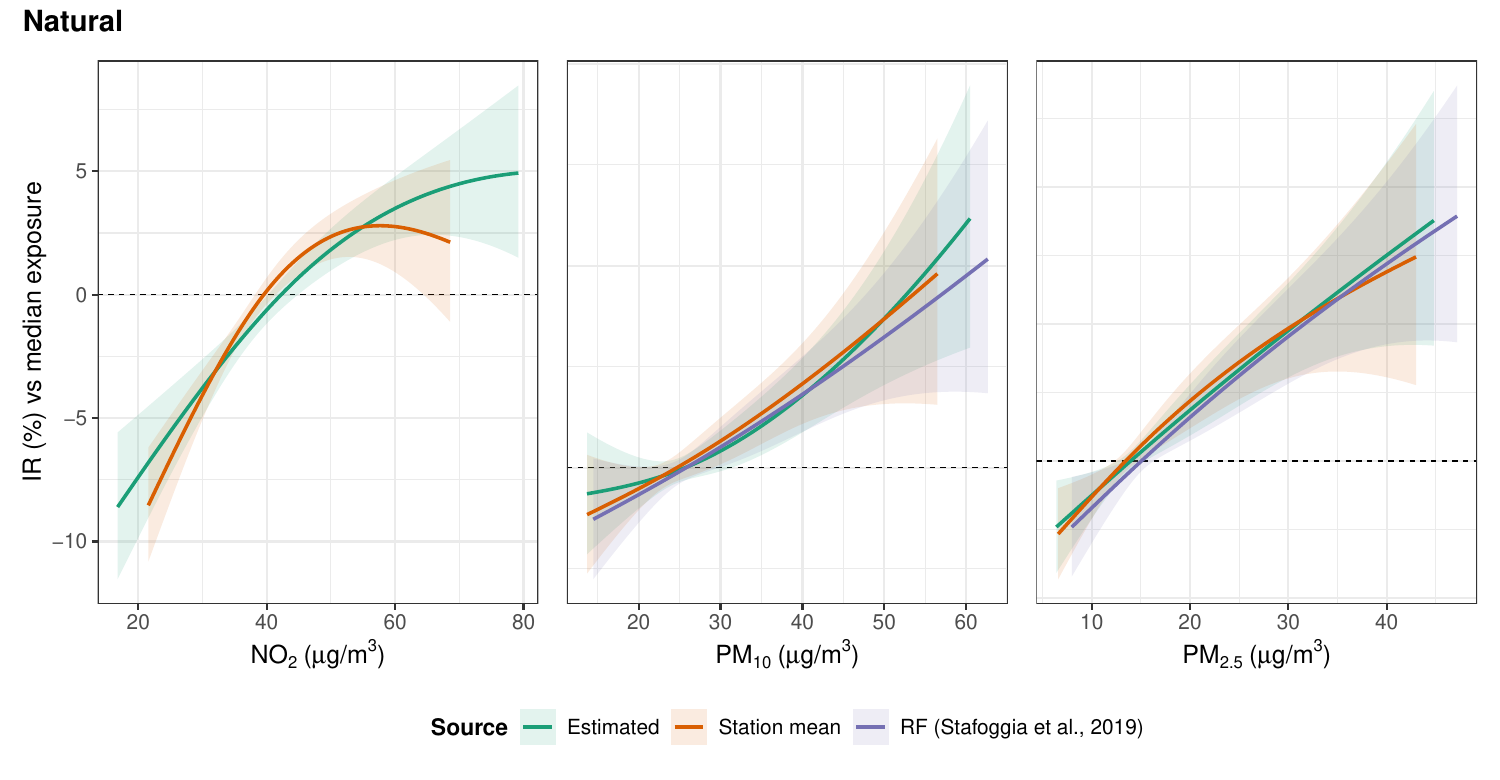}
  \caption{
  Exposure-response curves for the association between short-term air pollution exposure (lag~0--5 mean) and natural-cause mortality, Rome 2012--2019, estimated via natural-spline (2 df) case-crossover models, relative to the median exposure of each source: NO\textsubscript{2} (left), PM\textsubscript{10} (centre), and PM\textsubscript{2.5} (right). Lines represent three exposure sources: model-estimated (spatio-temporal Bayesian model), station mean (average of Rome monitoring stations), and random forest (RF) estimates from \citet{stafoggia_2019} (PM\textsubscript{10} and PM\textsubscript{2.5} only). Shaded areas indicate 95\% confidence intervals. IR (\%) = percent increase in risk relative to the median exposure.}
  \label{plot:curve_expresp}
\end{figure}

Also for this reason, in Figure~\ref{plot:curve_expresp} we report exposure-response curves using the natural-spline specification, for each pollutant. The three sources showed mostly concordant behaviour overall. NO\textsubscript{2} suggested a non-linear effect, as confirmed by the LRT: the model-estimated (green) line appears to grow until a plateau (around 80 $\mu$g/m$^3$), while the station-mean (orange) line appears to decline in risk from around 55 $\mu$g/m$^3$.
PM\textsubscript{10} showed a more linear pattern: curves increase over the whole domain, with general agreement among the three exposure sources.
Similar considerations hold for PM\textsubscript{2.5}, which showed the closest concordance among the three sources. In general, for all pollutants, the spatio-temporal model captured higher variability than the station mean (i.e. range of the green line is wider than that of the orange one).

We refer to the Supporting Information for a detailed interpretation of similar outputs for cardiovascular and respiratory causes. In general, the overall shape of the exposure-response relationship was similar across the three causes of death analysed, and consistent in behaviour across the data sources.

\clearpage

\section{Discussion}\label{EH::discussion}
Estimating the association between air pollution and its effects on human health still represents a huge epidemiological challenge. We proposed a Bayesian autoregressive spatio-temporal model, able to predict daily air pollution at fine grid scale (1~km) over the whole Rome municipality, and linked model-estimated exposures to geolocated mortality data through a case-crossover approach. For mean lag~0--5 exposure, results showed a significant positive association between each pollutant (NO\textsubscript{2}, PM\textsubscript{10}, and PM\textsubscript{2.5}) and natural-cause mortality. As an additional validation check, a sensitivity comparison against observed station data and already-validated random-forest estimates confirmed high concordance across the three exposure sources, both in magnitude and in shape of the estimated exposure-response relationships. We interpret these results as evidence that the fitted models show good predictive performance on average, and can lead to estimated associations in terms of human health effects comparable to other validated sources. \\

Both the exposure model and case-crossover results help in explaining some of the observed patterns. Fixed-effect estimates showed limited discriminative power for all three pollutants (only the weekend/holiday indicator had a 95\% CrI excluding zero, for NO\textsubscript{2} and PM\textsubscript{10}, suggesting generally lower concentrations during non-working days). Sensitivity checks (Supporting Information), however, confirmed that the inclusion of fixed effects did not affect predictive performance: we decided to retain them in the final specification given their established relevance in air-pollution modelling. Furthermore, estimated exposure surfaces reflected spatial behaviours of each pollutant: maps for PM\textsubscript{2.5} were overall smoother, due both to the lower number of stations available (only 8, compared to 12/13), and because of the more spatially homogeneous distribution of this pollutant. Some spatial patterns in the estimated surfaces could be explained by considering monitoring-network characteristics together with established differences in pollutant behaviour: for instance, the PM\textsubscript{10} peak near the Tiburtina station is likely attributable to the absence of a nearby background station monitoring the same pollutant (Preneste monitors only NO\textsubscript{2}). This is also reflected by the estimated decay parameter: even if the values are not directly comparable between pollutants due to different response transformations, the estimate for NO\textsubscript{2} was about one order of magnitude larger than those for the other pollutants, suggesting a shorter effective spatial range, more driven by traffic dynamics and micro-scale variability, compared to a higher smoothness typical of particulate matter.
Moving to the epidemiological results, the observed difference for NO\textsubscript{2} between the model-estimated and station-mean exposure-response curves at high concentrations could reflect models' limited ability to capture the extremes of the distribution (Supporting Information), especially on the right tail. Moreover, the results of the proposed workflow support the importance of evaluating the exposure model in two distinct ways. While cross-validation confirmed good statistical performance in a leave-one-site-out setting, the case-crossover sensitivity comparison showed that model-estimated exposure produces mortality associations in line with those obtained from independent sources. \\

These findings extend existing evidence on air pollution and mortality in Rome and in Italian urban contexts more generally. An early time-series study already reported a significant short-term association between air pollution and daily mortality in the city \citep{michelozzi1998}, later corroborated by a large cohort study linking long-term exposure to particulate matter and nitrogen dioxide with cause-specific mortality in the Rome population \citep{cesaroni2013}. More directly comparable to the present analysis, a time-series study covering non-accidental mortality in Rome over 1998--2014 estimated, for mean lag~0--5 exposure, incidence rate increases of 1.46\% for PM\textsubscript{10}, 1.75\% for PM\textsubscript{2.5}, and 3.03\% for NO\textsubscript{2} per 10~$\mu$g/m\textsuperscript{3} increase \citep{Renzi2017}, of comparable magnitude to the associations estimated here (1.3\%, 2.4\%, and 2.1\%, respectively). Similar short-term associations have also been reported in other major European capitals, from traffic-specific pollution markers and cardiovascular/respiratory mortality in London \citep{Atkinson2016}, to particulate matter (also including the effect of the desert dust) and cause-specific mortality in Madrid, the latter estimated through the same time-stratified case-crossover design adopted here \citep{Diaz2012}.\\

Some limitations of the work should be acknowledged. First, the models were estimated using a sparse sample of monitoring stations, confined to a relatively small spatial range within the centre of the Rome municipality (minimum inter-station distance 2.7~km, maximum 33~km), and were used to generate predictions across all 1,292 1$\times$1~km grid-cells. This, together with the fact that model behaviour was mainly driven by the spatio-temporal random effect (leaving almost no space for fixed effects informative contribution, as already discussed), means that inference in places far from monitoring sites (i.e., far from the centre of the municipality) should be interpreted with additional caution. On the other hand, however, it is expected that most of the population resides in areas well represented by the monitoring stations, making the model estimates a good surrogate for population exposure. The limited number of monitoring sites, therefore, shaped both the structure and the predictive behaviour of the models: this also likely explains why cross-validation showed greater difficulty in capturing high-traffic contexts, especially for NO\textsubscript{2}. A related limitation is the difficulty of the proposed models in capturing the tails of the distribution: as also discussed in the Supporting Information, model assumptions hold mainly for the bulk of the distribution, leaving greater uncertainty about their validity in the left and right tails. Furthermore, we did not adjust for other pollutants in each time-stratified regression, leaving an essentially univariate analysis despite the well-established inter-relationships among pollutants, particularly in urban settings. Finally, we also did not explore stratifications by sex or age across causes, which could reveal additional heterogeneity in the estimated associations.\\
Two future developments of this work are also worth discussing, among several possible. First, the proposed Bayesian spatio-temporal framework could be extended to the right tail of the pollutant distribution, formalising it through statistical models for extremes, such as quantile regression, to better capture the health effects of extreme pollution episodes. Second, moving to larger spatial scales such as regional or national where usually a larger monitoring network is available, could offer an opportunity to improve the characterisation of meteorological and climatic fixed effects: this might improve their discriminative power, potentially through spatially-varying coefficients.\\

This study confirms a significant positive association between short-term exposure to NO\textsubscript{2}, PM\textsubscript{10}, and PM\textsubscript{2.5} and natural-cause mortality in the Rome population, based on a validated Bayesian spatio-temporal exposure model. Together with the sensitivity comparison using different exposure sources, the results reinforce the value of fine-scale, spatially-resolved exposure modelling for urban air-pollution epidemiology. Although Rome, like other major European cities, has introduced new traffic- and emission-related regulations in recent years, air pollution remains an important public-health concern. The validated workflow proposed in this work offers a transferable framework for cities facing similar constraints in monitoring-network density.

\section*{Associated Content}
\subsection*{Data Availability Statement}
The dataset used in this study for exposure assessment is completely derived from publicly available sources. Derived exposure data and the corresponding code will be made available in a public repository upon acceptance. Cohort mortality data provided by Lazio Region Health Department/ASL Roma 1 cannot be made publicly available. Code for the case-crossover analysis will also be made available in the same repository, though it requires access to the restricted cohort dataset to be executed.
\subsection*{Supporting Information}
The following Supporting Information is available free of charge, provided as a separate document:
\begin{itemize}
\item Supporting Information: Additional descriptive analyses of exposure and mortality data, model selection and cross-validation results, residual diagnostics, and cardiovascular/respiratory case-crossover results (PDF).
\end{itemize}

\section*{Author Information}
\subsection*{Author Contributions}
ER collected and curated the data, performed the analyses, created figures and tables, and wrote the manuscript. MS and GJL contributed to the statistical analysis and interpretation of results. MS and PM contributed to data acquisition and study design. GJL and MS conceived, supervised the study, and contributed to the interpretation of findings. All authors reviewed and approved the final manuscript.
\subsection*{Notes}
The authors declare no competing financial interest.
\subsection*{Ethics approval and consent to participate}
Not applicable. This study used administrative mortality data with no direct contact with study subjects.

\section*{Funding}
Project implemented with the technical and financial support of the Ministry of Health -- PNC PREV-A-2022-12376. Additional support was provided by the Department of Statistical Sciences, Sapienza University of Rome, project \textit{Leaving No One Behind: Methods for Sustainability} (ID nr.: RD124190DA1146AA - CUP: B83C24007080005). 

\section*{Acknowledgments}
Work carried out thanks to the support of the Air and Health Atlas Study Group. We are grateful to Emiliano Ceccarelli (Department of Statistical Sciences, Sapienza University of Rome; Rome, Italy) for his contribution to the development and implementation of the kriging algorithm. We also thank Federica Nobile (Department of Epidemiology, Lazio Region Health Service/ASL Roma 1; Rome, Italy) for data extraction and preprocessing of the cohort dataset. Finally, we are grateful to Jorge Castillo-Mateo (Department of Statistical Methods, University of Zaragoza; Zaragoza, Spain) for his helpful suggestions on the analytical approach adopted in this work.

\bibliography{RomaEpimean}

\end{document}